\documentclass[aps,prd,preprint,groupedaddress]{revtex4-1}
\usepackage[utf8]{inputenc}
\usepackage{amsmath, amsfonts, mathtext, amssymb, mathrsfs,subfigure,wrapfig}
\input epsf
\usepackage[utf8]{inputenc}
\usepackage{graphicx} 

\usepackage{textcomp}
\newcommand*{\No}{\textnumero} 

\begin{document}

\title{Horndeski black hole observational properties}

\author{D.A. Tretyakova}
\email[]{daria.tretiakova@urfu.ru}
\affiliation{Institute for Natural Sciences, Ural Federal University, Lenin av. 51, Yekaterinburg, 620083, Russia}

\date{\today}

\begin{abstract}
We examine geodesics for  scalar-tensor black holes in the Horndeski-Galileon non-minimal kinetic coupling framework. Our analysis shows that bound orbits may not be present within some  model parameters range. Using the observational data we pose bounds on the solution parameter values, as well as  model parameters.
\end{abstract}


\maketitle

\section{Introduction}
Recent observations indicate that GR might indeed be modified at cosmological distances: our Universe experiences an accelerating phase of expansion \cite{2013ApJS19H}. A possible interpretation of this expansion in terms of General Relativity (GR) states that about $70\%$ of the total energy of our Universe is attributed to the dark energy  with large and negative pressure, for which the cosmological constant $\Lambda$ is considered as the best fit nowadays (see \cite{Bamba:2012cp} for a comprehensive review on dark energy or \cite{2013FrPhyL} for a brief one). However the recent result of \cite{2013ApJS19H}, stating that the equation of state parameter for the dark energy is $w_{DE_0} = -1.17(+0.13 -0.12)$ for the flat Universe at the $68\%$ C.L., disagrees with purely cosmological constant dark energy $w_{DE_\Lambda}=-1$.  This may be considered as an indication of the dynamical (at least in part) dark energy. In any case the origin of the late-time acceleration remains unknown.

Scalar-tensor gravity is a widely accepted alternative to the General Relativity. The most general scalar-tensor  action resulting in the second order field equations was  proposed by Horndeski \cite{Horndeski:1974wa}. The same result was rediscovered by studying the covariant Galileons \cite{Deffayet:2009mn, Deffayet:2009wt},  a ghost free scalar effective field theory containing higher derivative terms that are protected by the Galileon shift symmetry. The action for the  Horndeski/Galileons scalar-tensor model  reads
\begin{equation}
S = \int dx^4 \sqrt{-g} \left( \zeta R - \eta \left( \partial \phi \right)^2 + \beta G^{\mu\nu} \partial_\mu \phi \partial_\nu \phi - 2\Lambda \right) \label{ac},
\end{equation}
here $ G^{\mu\nu} $ is the Einstein tensor, $ \phi $ is the scalar field,  $ \zeta>0, \eta$ and $\beta$ are model parameters. Though the Horndeski non-trivial kinetic coupling sector is not exhausted by this action we will restrict our consideration to the action above, since most of the static spherically symmetric solutions we are about to consider relate to this framework\footnote{Note also that the second component of the non-trivial kinetic coupling sector, the so-called Paul term, experiences problems with  describing  neutron
stars \cite{Maselli:2016gxk}.}.  
The model \eqref{ac} is known to admit a rich spectrum of cosmological solutions (see \cite{Starobinsky:2016kua} and references therein) describing the late-time acceleration and the inflationary phase. Moreover, for $\eta\neq 0$ it admits solutions for which the $\Lambda$-term is totally screened, while the metric is not flat but rather de Sitter with the Hubble rate proportional to $\eta/\beta$. It offers an exciting opportunity to describe the late time cosmic self-acceleration while screening the vacuum $\Lambda$-term and hence circumventing the cosmological constant problem. Paper \cite{Kimura:2011qn} shows that  purely kinetic coupled gravity is inconsistent with the constraint from the gravitational Cherenkov radiation for any theoretically allowed parameter $\beta$. Introducing $\Lambda$ in the action \eqref{ac} (dividing the dark energy into different-behaving components)  is a way to relax this inconsistency. Henceforth it would be interesting to pursue the study of the solutions of \eqref{ac} at the astrophysical  and solar system scale, which implies static and spherically symmetric solutions.

This model can be integrated completely in the static and spherically symmetric sector \cite{Charmousis:2015aya} and numerous black hole solutions are presented in the literature. The key ingredient of this solutions is a scalar field linear time dependence, which seems like a natural feature on a cosmological background. Although the Vainstein screening mechanism is generally at work in Horndeski gravity, in the case of a minimal coupling of the scalar field to matter no screening radius can be posed. The solution, considered as a candidate to represent astrophysical objects must then posses de Sitter asymptotics\footnote{The asymptotic structure of the solutions might also be interesting in the context of AdS/CFT correspondence and in brane cosmology, along the lines of \cite{Birmingham:2001dq}.}.

Many of the black hole solutions of \eqref{ac} posses similar properties, so we try to perform general analysis, allowing to reduce the number of viable solutions. Such a reduction meets the needs, since in extended gravity we are usually left aside of the Birkhoff's theorem and forced to choose between multiple vacuum solutions. Reducing the parameter space and hence the number of solutions available could  then help to confront the theory with observations.

An  example, demonstrating that studying local spherically symmetric solutions can provide useful information on cosmological models is given in \cite{Tretyakova:2015vaa}. The paper  implements some PPN-based bounds on the model parameters due to the black hole-like metric for the framework, initially constructed for cosmological purposes. The parameter region left does not satisfy the purposes of the model anymore, rendering it far less attractive. Therefore, weak field and  geodesic analysis are useful tools to explore extended gravity solutions, even though the observable effects are small or the current observations may not be sensitive enough to distinguish extended gravity from the GR.

Any modification of GR must be consistent with  astrophysical and Solar  System  scale constraints,  which  are  very  stringent. The Schwarzschild solution in GR also describes the exterior of any spherically symmetric body in the weak field limit (hence the Solar System), and so must do it's analog in extended gravity theory. The purpose of this paper is thus to threat the local  spherically symmetric solutions of  \eqref{ac}  as astronomical objects and see if this picture is in agreement with the observed one. To do so we study the test particle  motion around the compact object described by the geodesic equations. For this sake we choose the metrics with de Sitter asymptotic behavior. 

This paper is organized as follows. In section \ref{s1} we briefly summarize the properties of the black hole-like solutions for the action \eqref{ac}. We also introduce a new  black hole-like solution and briefly discuss it's properties. Section \ref{s2} contains solution's parameter estimates found in the literature as well as some new ones, obtained in this paper. We proceed with analyzing geodesic equations in section \ref{s3}. Conclusions are given thereafter.

\section{Horndeski/Galileon black hole} \label{s1}
For the spherically-symmetric ansatz
\begin{equation}
ds^2 = h(r)dt^2 - \frac{dr^2}{f(r)} - r^2d \Omega^2
\label{ds^2}
\end{equation}
black hole solutions were found in the series of papers \cite{Rinaldi:2012vy}-\nocite{Babichev:2013cya, Charmousis:2015aya, Babichev:2015rva, Anabalon:2013oea}\cite{Minamitsuji:2013ura}. These solutions possess very similar properties, all governed by the master equations: 
\begin{eqnarray}
&& f(r)  =  \dfrac{( \beta + \eta r^2) h(r)}{\beta \left( rh(r)\right)'},\label{f0} \\
&& h(r)  =  -\frac{\mu}{r} + \frac{1}{r} \int\dfrac{k(r)}{\left( \beta + \eta r^2\right)}dr,\\
&&\phi(r)  =  qt+\psi(r), \label{phi}
\end{eqnarray}
where $ \mu$ plays the role of the mass term and $k$ should be derived by means of the following constraint equation:
\begin{eqnarray}
&& q^2\beta\left( \beta + \eta r^2\right)^2 -
\left( 2\zeta\beta + \left( 2\zeta\eta - \lambda\right) r^2\right)k + C_0 k^{\frac{3}{2}} =0. \label{k}
\end{eqnarray}
Here $C_0$ is an integration constant. By introducing a mild linear dependence in the time coordinate for the scalar field  one evades the scalar field being singular for it's derivative on the horizon  \cite{Charmousis:2014zaa} and makes the field equations to bifurcate the no-hair theorem at the same time.  The shift symmetry is keeping the field equations time-independent and consistent with the static ansatz. This permits asymptotically flat (or de-Sitter) solutions and crucially gives regular scalar tensor black holes. By using different parameter combinations the equations above can be integrated to give various solutions, among which is the stealth de Sitter one 
\begin{eqnarray}
f(r) & = & h(r)=1-\cfrac{\mu}{r} +\cfrac{\eta}{3\beta}r^2 , \label{f1}\\
q^2&=&(\zeta\eta+\beta\Lambda)/(\beta\eta), C_0=(\zeta\eta-\beta\Lambda)\sqrt{\beta}/\eta.
\end{eqnarray}

In the paper \cite{Charmousis:2015aya} the authors show that the de Sitter Schwarzschild solution is not an isolated solution, instead, it is continuously related to a full branch of de Sitter like black hole solutions with similar characteristics. The authors support this statement by obtaining a solution, deviating slightly from the de Sitter one and having the form 
\begin{eqnarray}
h(r) & = & C-\frac{\mu}{r} +Ar^2 +\Delta, \qquad \Delta=B\dfrac{\arctan (r \gamma)}{r \gamma}, \label{h_gen}
\end{eqnarray}
which is asymptotically de Sitter for $A<0$. The coefficients read
\begin{eqnarray}
A & = & -\cfrac{\eta}{3|\beta|}, B= \cfrac{2(1+\gamma^2)\epsilon}{\zeta+y}, \epsilon <<|y-1|, \nonumber\\
\gamma&=& \sqrt{\cfrac{\eta}{|\beta|}}\cfrac{\zeta+y}{\zeta-3y}, C=1-\cfrac{2\epsilon}{\zeta+y}, y=\cfrac{\Lambda|\beta|}{\eta}, \label{apds}
\end{eqnarray}
where $\epsilon$ is a small parameter, marking the deviation of the solution from the stealth de Sitter one. 

By examining the known solutions we see that \eqref{h_gen} is a very common expression, joining many particular solutions (differing by the parameters $A,B,C,\gamma$). Furthermore, many other solutions of a kind can be easily constructed. 
Indeed,  for the values $ C_{0} = 0 $ and $ q \neq 0 $ the master equations can be integrated to give
\begin{eqnarray}
	&& A=0, C = \dfrac{q^2 \eta}{2 \zeta \gamma^2}, \quad
	 B = \dfrac{q^2 \beta}{2 \zeta } \left( 1 - \dfrac{\eta}{\beta\gamma^2} \right), \quad \gamma = \sqrt{\dfrac{\eta}{2\beta} - \dfrac{\Lambda}{2\zeta}}.
\end{eqnarray}
This solution is new, though very similar to those previously known, asymptotically  equivalent to the black hole in the Einstein static universe. So a class of solutions governed by \eqref{h_gen} is vast and not limited to asymptotically de Sitter spacetimes. 

Many crucial space-time properties are specifically related $h(r)$ and  by using  this metric function we can pose  bounds on a wide class of solutions. The most of the bounds we will obtain rely on the observational data, so they would apply to the solutions possessing correct asymptotic behavior. Such explicitly written down solutions are nowadays \eqref{f1} and \eqref{apds} displayed above. However new solutions of this branch can be  constructed along the guidelines presented in \cite{Charmousis:2015aya}. The goal of this paper is to threat this solutions as a class and withdraw some general conclusions. So in what follows we will use the metric in the form \eqref{h_gen}, assuming that the results apply to any solution of that kind.

\section{Parameter estimates} \label{s2}


Let us first briefly review the constraints on the parameters of the non-minimal derivative coupling sector of Horndeski theory, determined elsewhere. The metrics from \cite{Babichev:2013cya} were widely discussed in the literature \cite{Cisterna:2016vdx,Cisterna:2015yla, Ogawa:2015pea}.
First of all, the effective gravitational constant for the action \eqref{ac} according to \cite{Kimura:2011dc} can be expressed as
\begin{eqnarray}
&& |\dot G / G|  =  \left | \cfrac{-3\beta \dot X}{\zeta +3\beta X} \right|, \quad X \equiv-\cfrac{1}{2}\partial_{\mu}\phi\partial^{\mu}\phi
\end{eqnarray}
where the dot denotes the time derivative. Since according to \eqref{phi} $\dot X =0$ we have $|\dot G / G|=0$ in perfect agreement with GR.

One should keep in mind that for the model in question the spin-0 degree of freedom also acquires dynamics via the kinetic mixing with the spin-2 graviton. Therefore the condition for the solution to be ghost-free does not just boil down to the ``right'' kinetic term sign. See e.g. \cite{Deffayet:2010qz}, where a Galileon model has been shown to be stable on cosmological solutions for the unusual sign of the standard kinetic term.


Minamitsuji \cite{Minamitsuji:2014hha} investigated the stability of BH solutions under massless scalar perturbations in the nonminimal derivative coupling subclass. The quasinormal modes can be computed, and no unstable modes were found. Considering the same BH solutions, Cisterna et al. \cite{Cisterna:2015uya} found that these black holes are stable under odd-parity gravitational perturbations as well. 
The stability conditions of hairy black holes in the non-minimal derivative coupling sector can be extracted from \cite{Ogawa:2015pea}, keeping in mind that  $X \neq Const$ due to the radial dependence of the scalar field.


Let us turn to the particular black hole spacetime. One of the simplest requirements that can be imposed on the black hole metric is the presence of an event horizon. For the Schwarzshild --- de Sitter spacetime this would require for the cubic $r h(r)$ of \eqref{f1} to have three real roots, which happens whenever $A \in \left(- 4/27\mu^2,0 \right)$. 
Applied to the SgrA*\footnote{$M_{Sgr A}=4\times 10^6 M_{\odot}$}  black hole this requirement would state $|\eta/\beta| < 4.5\times 10^{-21}$.

Another demand is that the metric should admit bound orbits, which may in general not be the case for a Schwarzshild --- de Sitter kind of spacetime. According to the results of \cite{Jaklitsch1989} this would correspond to $|\eta/\beta|<  16/1875\mu^2$ or for SgrA* to $|\eta/\beta|<  8.5 \times 10^{-23}$.

We can also consider weak-field observations. One of the well-studied gravitational effects is the frequency shift for the satellite on the Earth orbit
\begin{equation}
\cfrac{\delta \nu}{\nu}=1-\sqrt{\cfrac{h(R+d)}{h(R)}}\approx \cfrac{V(R+d)-V(R)}{c^2},
\end{equation}
up to the first order in the weak field approximation with $d$ being the satellite orbit height and $R$ - the Earth surface radius, $V$ - the corresponding gravitational potential. For the metric \eqref{h_gen} there should be an additional shift, related to the deviation of h(r) from the Schwarzshild solution
\begin{eqnarray}
&& 2\cfrac{\delta \nu}{\nu}\approx \delta_{Schw}+\delta_1+\delta_2 =\nonumber \\
&& \qquad \left ( \cfrac{\mu}{R}-\cfrac{\mu}{R+d} \right)+ A( (R+d)^2-R^2) + \left ( \Delta(R+d)-\Delta(R) \right).
\end{eqnarray}
Modern frequency measurements are in agreement with GR, hence we can make bounds using the frequency measurement accuracy $10^{-14}$ achieved in the GP–A redshift experiment\footnote{$d= 15\times 10^{3} km, M_{\oplus}=5,972\times 10^{24} kg, R_{\oplus}=6371 km, \\ G_0=6.67384 \times 10^{-11} m^3 kg^{-1} c{-2}$} \cite{PhysRevLett.45.2081}. Numerical estimates show that $\delta_1$  does not exceed the accuracy of the relative frequency measurement when
\begin{eqnarray}
A&<&2.4\times 10^{-29}.
\end{eqnarray}
This estimate allows us to set $A\approx 0$ when considering accretion and null geodesics for the Sgr A* black hole or any smaller one since the conditions $Ar^2<<\mu/r$ and $Ar^2<<\Delta$ are well satisfied  within the corresponding  $100r_{Schw}$ distance. This agrees well with the fact that the expansion of the universe is not known to manifest itself in the Solar System. So, we may neglect the de Sitter term on astronomical scales and the metric can be considered in the form
\begin{eqnarray}
h(r) & =& C-\frac{\mu}{r} +\Delta \label{h0}.
\end{eqnarray}

Let's see how this bounds would affect the known solutions. Take the approximately Schwarzshild - de Sitter solution \eqref{apds} as an example. The solution is valid for $y/\zeta\in (-1,1/3)$. The second multiplier in $\gamma$ is of order unity except the special case $\zeta\approx 3y$. Hence, excluding this fine tuned case we can see that $\gamma \sim \sqrt{|A|}$. So we can further suggest that due to the  smallness of $\gamma$, $\arctan(r\gamma)\approx r\gamma$ and hence
\begin{eqnarray}
\Delta&\approx&B, \\
h(r) & =& (C+B)-\frac{\mu}{r}. \label{h1}
\end{eqnarray}
This is a spacetime of a black hole with a global monopole (up to the difference in f(r)).  This kind of black hole  was previously studied  in the literature for flat and de Sitter spacetime, revealing the expressions for the deflection angle \cite{Shi:2009nz}, perihelion precession \cite{Hao:2003rz} and accretion disc radiant energy flux \cite{1674-1056-23-6-060401}. Note that the global monopole black hole metric displays the solid angle deficit. We further use the equation \eqref{f0} to define $f$ with respect to $\eqref{h_gen}$ and neglect terms $\sim Ar^2$ to keep our approximation. This would imply
\begin{equation}
f(r)\approx\cfrac{h(r)}{B+C}. 
\end{equation}
For the global monopole case this would mean that the area of the sphere is restored while $h$ is ambiguous by the constant time rescaling.  We will further use \eqref{h0} as well as \eqref{h1} to study geodesic motion and withdraw some conclusions.

Just to confirm again that the metric structure \eqref{h_gen} of the solution above is a common one and does generally not rely the smallness of $\epsilon$, consider the solution from \cite{Babichev:2013cya}:
\begin{eqnarray}
C&=&1 , A =  \dfrac{\gamma^2}{3}\dfrac{\zeta - y}{3\zeta + y} ,  B =  \dfrac{(\zeta+y)^2}{4 \zeta^2 - (\zeta+y)^2}, \gamma = \sqrt{\eta /3\beta},  \label{bab}
\end{eqnarray}
where obviously $\gamma \sim \sqrt{A}$ analogously to the case above. Hence the approximations made due to the constraint above are of general order. 

\section{Geodesic motion} \label{s3}
Geodesic equation  for the radial coordinate reads
\begin{eqnarray} \label{geodesic}
&& \left( \frac{dr}{d\varphi} \right)^2 = f(r)P(r), \\
 && P(r)=\left[ \cfrac{E^2-jh(r)}{h(r)}\right] \cfrac{r^4}{L^2}  -r^2,
\end{eqnarray}
where $j=0$ for massless particles and $j=1$ for the massive ones, $E$ and $L$ are the energy and momentum of the test particle per it's unit mass respectively. We will further use the inverse radius $ u = r^{-1}$ as it is usually done. Working with geodesics we are beyond the black hole horizon and hence $h(u)\neq 0$ (and $f(u)\neq0$). Therefore we can write  \eqref{geodesic} for the circular orbits as
\begin{eqnarray}
 P(u) &=& 0, \label{cir1} \\
P'(u) &=& 0. \label{cir2}
\end{eqnarray}
We set $E^2<1$ for the bound orbits and the prime denotes the derivative with respect to the radial coordinate. Note that the circular orbits are thus determined solely by $h(r)$.  Substituting \eqref{h_gen} into \eqref{h0} we obtain 
\begin{eqnarray}
&&  \mu u^3 - Cu^2 +\dfrac{\mu}{L^2} u - \dfrac{C-E^2}{L^2}   -\Delta \left(u^2+ \dfrac{1}{ L^2}\right) =0, \label{P}\\
&& \!\!\!\! 3\mu u^2  - 2Cu +\dfrac{\mu}{L^2}    -\Delta' \left(u^2+ \dfrac{1}{ L^2}\right) - 2u \Delta  =0. \label{P'}
\end{eqnarray}
One can simplify $ \Delta' $ if we recall that $ \gamma^2/u^2 \equiv \gamma^2 r^2 \sim Ar^2 $, and neglect this term.
We now extract  $ \Delta$ from \eqref{P} and substitute into \eqref{P'}:
\begin{equation}
\cfrac{-(B+C)L^4u^4+(3E^2-2(B+c))L^2u^2+E^2-(B+C)}{(3L^2u^2+1)L^2}=0 \label{F(u)}
\end{equation}
 Solving the above results in the following expression for the circular geodesics
\begin{eqnarray}
 u_{\pm}&=& \cfrac{\sqrt{2(3\tilde{E}^2-2 \pm 2\sqrt{9\tilde{E}^4-8\tilde{E}^2})}}{2L}, \label{u_cir}\\
\tilde{E}&=&E/\sqrt{B+C}
\end{eqnarray}
The plus sign denotes the stable orbit and minus --- the unstable one, corresponding to the stationary points of the effective potential.  Note that $A\neq0$ would rise the order of \eqref{F(u)} to six, adding one root being just one more unstable circular orbit. We see that circular orbits for massive particles are available when $E^2\geq 8(B+C)/9$.
But the bound circular orbit has to obey $E^2<1$, hence 
\begin{equation}
0<(B+C)<\cfrac{9}{8}. \label{circon}
\end{equation}
At this point we would like to digress for a moment to demonstrate how  the results obtained above could be applied.
Circular orbits represent a narrow subclass of relativistic bound orbits. But with circular orbits being forbidden, all of the bound orbits are  eliminated. This leaves us only with the trajectories, ending up in the black hole singularity, which would be a highly pathological behavior contradicting to the black hole accretion  \cite{2011ApJ...742...67M,2012Sci...337..540F} and Solar System observations. This means that the requirement \eqref{circon} can be used as a criterion to pick viable solutions (or at least parameter ranges). Consider (just for example) the metric \eqref{bab}.  For a negligible value of $\Lambda$ we can immediately obtain $B=1/3$ and  circular geodesics would be impossible, \eqref{circon} requires $\Lambda<0$, approximately $\Lambda<-17\zeta\gamma/2$. This is an interesting feature since $q=0$ for this metric. This might indicate that $q\neq 0 $ might yet be a crucial feature to get healthy black holes. 

\begin{figure}
\center{\includegraphics[width=0.9\linewidth]{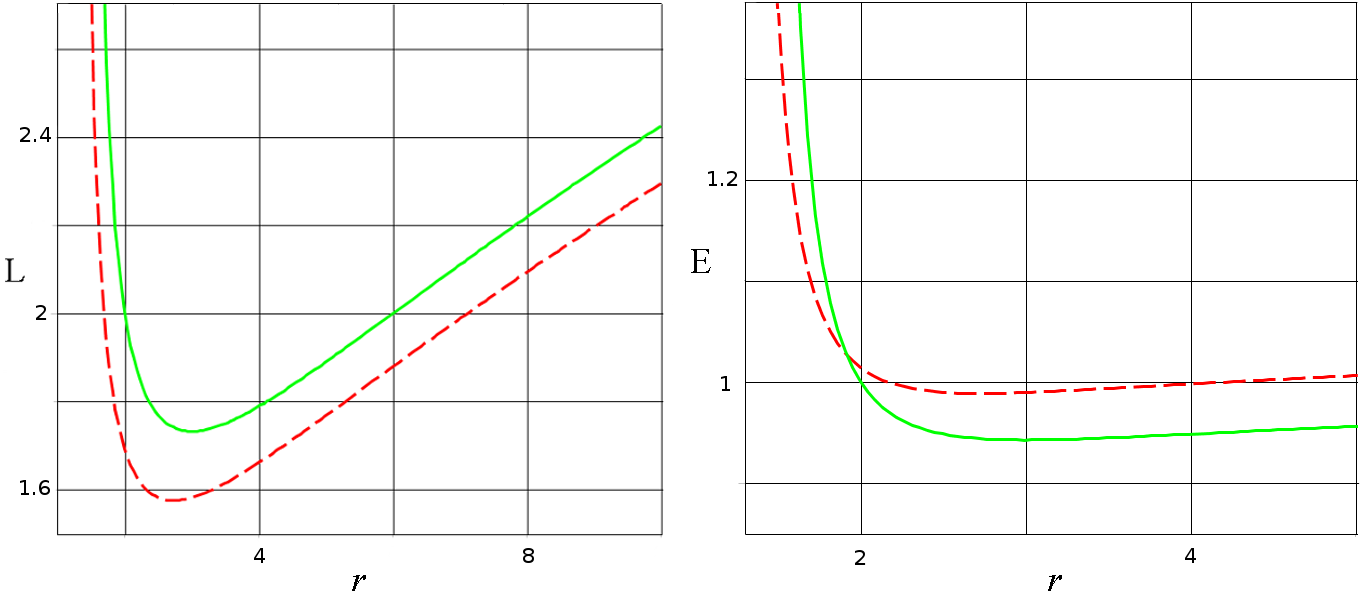}}
\caption{{Energy and angular momentum for the test particle on the circular orbit with $\mu=1,C=1,\gamma=10^{-14}$. The solid line corresponds to $B=0$, the dashed one to $B=1/10$. The allowed energy range squeezes as $(B+C)$ grows.}}
\label{EL}
\end{figure}

\begin{figure}
\center{\includegraphics[width=0.9\linewidth]{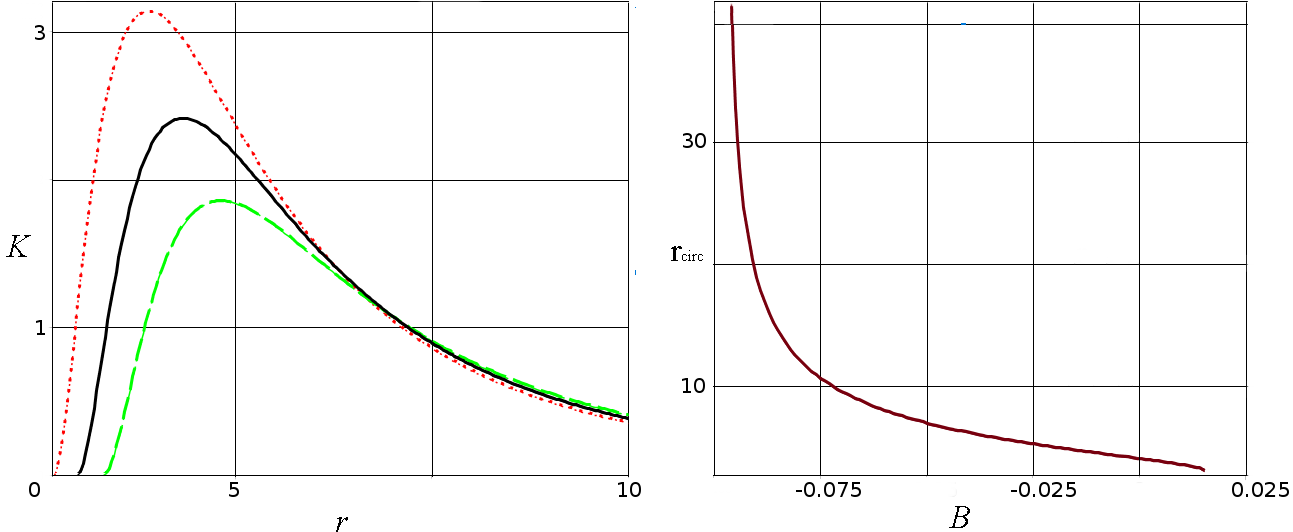}}
\caption{{\it{Left:} Radiant flux of an accretion disc surrounding a black hole$\mu=1,C=1,\gamma=10^{-14}$, free parameter counts $B=1/9, 0, -0.1$ upside down. \it{Right:} The radius $r$ of the circular orbit for the fixed $E$ and $L$ vs. $B$.}}
\label{fr}
\end{figure}

We now proceed with circular orbits. Whereas \eqref{u_cir} looks like the classical expression for the circular orbit, the  energy and momentum 
\begin{eqnarray}
L^2&=&\cfrac{\mu u+B-\Delta(u)}{u^2(-3\mu u-B+2C+3\Delta(u))}, \label{e}\\
E^2&=&\cfrac{2h(u)^2}{-3\mu u-B+2C+3\Delta(u))}, \label{l}
\end{eqnarray}
differ for the same radius, as one can see from the fig. \ref{EL}. The corresponding energy on the circular orbit grows with $(B+C)$, while the moment decreases. One  can also see from the fig. \ref{EL} that the allowed energy range squeezes as $(B+C)$ grows. The radius $r$ of the circular orbit for the same $E$ and $L$ would decrease with rising $B+C$ and vice versa (see fig. \ref{fr}). 
The innermost stable circular orbit can be evaluated exactly for the global monopole approximation giving\footnote{All expressions reduce to the Schwarzshild ones for $B=0,C=1$.}
\begin{equation}
E^2_{ISCO}=\cfrac{8(B+C)}{9}, \quad u_{ISCO}=\cfrac{(B+C)}{3\mu}, \quad L_{ISCO}^2=\cfrac{3\mu^2}{(B+C)^2}.
\end{equation}
All the above could potentially affect the observable characteristics of a black hole entourage. 

In astrophysics the particles, orbiting a black hole can be detected  due to the emitted radiation when they form an accretion disc. Consider the accretion disc made from particles moving on circular orbits, spiraling on the black hole very slowly. The energy flux of such a disc reads \cite{1974ApJ...191..499P}
\begin{eqnarray}
 &&K=-\cfrac{M_0}{4\pi\sqrt{-det g}} \cfrac{\omega_{,r}}{(E-\omega L)^2}\int_{r_{ISCO}}^{r}(E-\omega L)L_{,r} dr,\\
&&\omega=\cfrac{\sqrt{\mu r+Br^2-\Delta(r)r^2}}{\sqrt{2}},
\end{eqnarray}
where $\omega$ is the angular velocity of the particle on the circular orbit and $E$ and $L$ are given by \eqref{e}-\eqref{l}.  
The result is plotted in the figure \ref{fr} for different values of $B+C$. The radiant flux is evidently affected by the correction terms, so that combining the data on the ISCO and the flux one could distinguish the given black hole (provided a good study of the accretion disc).
For the global monopole approximation the analytical expression for the flux can be found in \cite{1674-1056-23-6-060401}, however the direct observational verification by means of the accretion disc properties is probably the matter of the distant future.

There are also results for the monopole approximation obtained elsewhere that can be reinterpreted  to pose some bounds on $(C+B)$.
Paper \cite{Hao:2003rz} contains the the light deflection angle formula for the global monopole  which in our case would be 
\begin{equation}
\delta \varphi\approx  \cfrac{2\mu}{(C+B)^{3/2}R_0}.
\end{equation}
Given the current measurement accuracy we could put bounds on $(C+B)$. The paper \cite{10875} based on a number of VLBI measurements of angular separations of strong quasistellar radio sources passing very close to the Sun, claims a good agreement of the observations with the GR. The reported relative error in the  deflection angle is  $6.2\times 10^{-4}$, which would imply
\begin{equation}
|1-(C+B)|<3\times 10^{-4}. \label{b0}
\end{equation}
Note that a test particle would also have a different perihelion precession in the global monopole spacetime, 
\begin{equation}
\delta \phi\approx \cfrac{3\pi\mu}{a(C+B)(1-e^2)}.
\end{equation}
However since the accuracy of the perihelion precession data is of order of $10^{-3}$, the corresponding bound would be one order of magnitude weaker than \eqref{b0}. The perihelion precession could  be used as an observational indicator in the future, when a detailed knowledge of the galaxy center enviroment would be available. The precise ephemerides of the S-stars could then provide a good test of the black hole nature.


\section{Discussion and conclusions}
Departing from GR we are left aside of the Birkhoff's theorem. Black hole solutions in non-minimal derivative coupling sector of Horndeski/Galoleon scalar-tensor gravity, a viable extended gravity theory candidate,  suffer under this ambiguity. In this situation we are forced to choose between multiple spherically symmetric vacua and therefore some hints on how to determine the realistic solution would come in handy. 

In this paper we analyzed the asymptotically de Sitter branch of static spherically symmetric solutions in Horndeski/Galileon non-minimal derivative coupling framework. Our analysis  revealed that their observational characteristics may differ from those of the GR Schwarzshild solution substantially. To suppress such deviations several bounds are in order. For $(B+C)<0$ or $(B+C)>9/8$ bound orbits could not occur in the metric of the form \eqref{h_gen}. However to match the observations of the gravitational light deflection and perihelion precession we must demand $1-(C+B)<3\times 10^{-4}$ which guarantees that the bounds orbits are allowed for  $(B+C)>0$. From the frequency shift measurements we extract the bound on the de Sitter term $|\eta/\beta|<  7.2 \times 10^{-29}$. The structure of the accretion disc will be altered as well, caused by the change in the innermost stable orbit, particle energy and momentum, and, henceforth, the radiant energy flux. This might be a useful indicator for the observations in the far future.

\section{Acknowledgements}
This work was supported by Russian Foundation for Basic Research via grant RFBR \No 16-02-00682. We also thank D.A. Melkoserov and B.N. Latosh for useful comments.

\bibliographystyle{unsrt}
\bibliography{mybib}

\end{document}